\documentclass[12pt]{article}
\usepackage{amsmath}
\usepackage{amssymb}

\newcommand{\at}{allowed transformations}
\newcommand{\At}{Allowed transformations}

\newcommand{\gkpe}{generalized KP equation}

\newcommand{\gen}[1]{\partial_{#1}}
\newcommand{\curl}[1]{ \{#1\} }

\newtheorem{thm}{Theorem}

\numberwithin{equation}{section}
\begin{document}

\title{\bf \vspace*{-1.5in}
Generalized Kadomtsev-Petviashvili equation with an infinite
dimensional symmetry algebra }

\author{F. \textsc{G{\"u}ng{\"o}r}\\
\small
\begin{tabular}{c}
Department of Mathematics\\
Faculty of Science and Letters\\
Istanbul Technical University\\
Maslak, Istanbul, Turkey\\
\texttt{gungorf@itu.edu.tr}
\end{tabular}
\and
P. \textsc{Winternitz}\\
\small
\begin{tabular}{c}
Centre de Recherches Math{\'e}matiques\\
Universit{\'e} de Montr{\'e}al\\
C. P.~6128, Succ.~Centre-ville\\
Montr{\'e}al QC H3C 3J7, Canada\\
\texttt{wintern@crm.umontreal.ca}\\
\end{tabular}}
\date{}
\maketitle


\abstract{A generalized Kadomtsev-Petviashvili equation,
describing water waves in oceans of varying depth, density and
vorticity is discussed. A priori, it involves 9 arbitrary
functions of one, or two variables. The conditions are determined
under which the equation allows an infinite dimensional symmetry
algebra. This algebra can involve up to three arbitrary functions
of time. It depends on precisely three such functions if and only
if it is completely integrable. }
\section{Introduction}
A generalized Kadomtsev-Petviashvili equation with variable
coefficients has been proposed some time ago \cite{David87,
David89}. The motivation was to describe water waves that
propagate in straits, or rivers, rather than on unbounded
surfaces, like oceans. This equation was derived from the basic
equations of hydrodynamics, namely the Euler equations in three
dimensions. The assumptions made were similar to those used when
the KP equation is derived for water waves, namely weak
nonlinearity, weak dispersion, small amplitudes and propagation in
one direction (the $x$ axis) with the waves weakly perturbed in
the $y$ direction \cite{Johnson97, Ablowitz81, Ablowitz91}.
However, the derivation allowed for variable depth, the presence
of boundaries and of vorticity.

The generalized KP equation thus provides a description of
surface waves in a more realistic situation than the KP itself.
The additional terms and the variable coefficients make it
possible to treat straits of varying width and depth, variable
density and to take vorticity into account.

It was shown in the second article \cite{David89} that in special
cases, corresponding to specific geophysical situations, the
\gkpe{} is integrable. It then allows for soliton solutions with
"horse-shoe" shaped wave crests. Such waves are indeed observed,
for instance issuing from straits opening up into seas or oceans
\cite{Fu84, Lacombe82}.

The purpose of this article is to study the group theoretical
properties of a \gkpe {} that is somewhat more general than the
one introduced earlier \cite{David87, David89}, namely
\begin{equation}\label{1.1}
\begin{split}
& [u_t+p(t)uu_{x}+q(t)u_{xxx}]_{x}+\sigma(y,t)u_{yy}+a(y,t)u_{y} \\
&+ b(y,t)u_{xy}+c(y,t)u_{xx}+e(y,t)u_{x}+f(y,t)u+h(y,t)=0.
\end{split}
\end{equation}
We assume that in some neighbourhood we have
\begin{equation}\label{1.2}
p(t)\ne 0,\quad q(t)\ne 0,\quad \sigma(y,t)\ne 0.
\end{equation}
The other functions in \eqref{1.1} are arbitrary. More
specifically, our aim is to determine the cases when eq.
\eqref{1.1} has an infinite-dimensional symmetry group. The
motivation for this is two-fold. On one hand, the existence of an
infinite-dimensional symmetry group makes it possible to use Lie
group theory to obtain large classes of solutions. On the other
hand, integrable equations in 2+1 dimensions typically have
Kac-Moody-Virasoro symmetry algebras involving several arbitrary functions
of time \cite{david1, Levi88, Champagne88, Winternitz88, Orlov97}. This is
true for all equations of the KP hierarchy \cite{Orlov97}.
Moreover, these Kac-Moody-Virasoro symmetries can be directly
extracted from a much larger set of symmetries that includes
higher symmetries as well as nonlocal ones \cite{Orlov97,
Orlov86}.

Eq. \eqref{1.1} with $(p=q=1, f=h=0)$ was studied from a
different point of view by Clarkson \cite{Clarkson90} who showed
that the equation has the Painlev\'e property \cite{Ablowitz78,
Weiss83, Conte99} if and only if it can be transformed by a point
transformation into the KP equation itself.

In Section 2 we introduce "\at{}" that take equations of the form
\eqref{1.1} into other equations of the same class. That is, they
may change the unspecified functions in eq. \eqref{1.1}, but not
introduce other terms, or dependence on other variables. The \at{}
are used to simplify eq. \eqref{1.1} and transform it into eq.
\eqref{canon} that we call the "canonical \gkpe{}" (CGKP
equation). In Section 3 we determine the general form of the
symmetry algebra of the CGKP equation and obtain the determining
equations for the symmetries. In Section 4 we establish the most
general conditions under which the CGKP equation is invariant
under arbitrary reparametrization of time. This means that the
symmetry algebra contains the Virasoso algebra as a subalgebra. We
show that this Virasoro algebra is present if and only if the CGKP
equation can be transformed into the Kadomtsev-Petviashvili
equation itself by a point transformation. Section 5 is devoted to
the case when the CGKP equation is invariant under a Kac-Moody
algebra, (but not under a Kac-Moody-Virasoro one). Some
conclusions are presented in Section 6.
\section{Allowed transformations and a canonical generalized
Kadomtsev-Petviashvili equation} By definition "\at{}" are point
transformations $(x,y,t,u)\to
(\tilde{x},\tilde{y},\tilde{t},\tilde{u})$ that take eq.
\eqref{1.1} into another equation of the same type. That is, the
transformed equation  will  be the same as eq. \eqref{1.1}, but
the arbitrary functions can be different. The typical features of
the equation are that the new functions $\tilde{p}(\tilde{t})$
and $\tilde{q}(\tilde{t})$ depend on $\tilde{t}$ alone, the
others on $\tilde{y}$ and $\tilde{t}$, but no $\tilde{x}$
dependence is introduced. The only $\tilde{t}$-derivative is
$\tilde{u}_{\tilde{x}\tilde{t}}$, the only nonlinear term is
$\tilde{p}(\tilde{t})(\tilde{u}\tilde{u}_{\tilde{x}})_{\tilde{x}}$
and the only derivative higher than a second order one is
$\tilde{q}(\tilde{t})\tilde{u}_{\tilde{x}\tilde{x}\tilde{x}\tilde{x}}$.
These conditions are very restrictive and they imply that the
"\at{}" have the form
\begin{equation}\label{2.1}
\begin{array}{ll}
& u(x,y,t)=R(t)\tilde{u}(\tilde{x},\tilde{y},\tilde{t})-
\displaystyle\frac{\dot{\alpha}}{\alpha p}x+S(y,t),\\[.3cm]
& \tilde{x}=\alpha(t)x+\beta(y,t),\quad \tilde{y}=Y(y,t),\quad
\tilde{t}=T(t),\\[.3cm]
& \alpha\ne 0,\quad  R\ne 0\quad Y_{y}\ne 0,\quad \dot{T}\ne
0,\quad \dot{\alpha}f(y,t)=0.
\end{array}
\end{equation}
The coefficients in the transformed equation satisfy

\begin{equation}\label{2.2}
\begin{aligned}
  \tilde p(\tilde t) &  = p(t)\frac{{R\alpha }}
{{\dot T}},\;\;\tilde q(\tilde t) = q(t)\frac{{\alpha ^3 }}
{{\dot T}}, \\
  \tilde \sigma (y,t) &  = \sigma (y,t)\frac{{Y_y^2 }}
{{\alpha \dot T}}, \\
  \tilde a(\tilde y,\tilde t) &  = \frac{1}
{{\alpha \dot T}}\{ a Y_y  + \sigma Y_{yy} \} , \\
  \tilde b(\tilde y,\tilde t) &  = \frac{1}
{{\alpha \dot T}}\{ (b\alpha  + 2\sigma \beta _y )Y_y  + \alpha Y_t \} , \\
  \tilde c(\tilde y,\tilde t) &  = \frac{1}
{{\alpha \dot T}}\{ c\alpha ^2  + \beta_t   \alpha  + p(t)S\alpha
^2  + \sigma \beta _y^2
+ b\alpha \beta _y \} , \\
  \tilde e(\tilde y,\tilde t) &  = \frac{1}
{{\alpha R\dot T}}\{ R\alpha e - R\dot \alpha  + \dot R\alpha  + aR\beta _y  + \sigma R\beta _{yy} \} , \\
  \tilde f(\tilde y,\tilde t) &  = \frac{1}
{{\alpha \dot T}}f, \\
  \tilde h(\tilde y,\tilde t) &  = \frac{1}
{{\alpha R\dot T}}\{ h - \frac{d}{dt}\left( {\frac{{\dot \alpha }}
{{\alpha p}}} \right)   + p\left( {\frac{{\dot \alpha }} {{\alpha
p}}} \right)^2  + \sigma S_{yy}  + aS_y  + fS - e\frac{{\dot
\alpha }}{{\alpha p}}\}.  \\
\end{aligned}
\end{equation}

We now choose the functions $R(t), T(t)$ and $Y(y,t)$  in eq.
\eqref{2.1} to satisfy

\begin{equation}\label{2.4}
\begin{gathered}
  \dot T(t) = q(t)\alpha ^3 (t),\quad R(t) = \frac{q}
{p}\alpha ^2,  \hfill \\
  Y_y  = \alpha ^2 \sqrt {\left| {\frac{{q(t)}}
{{\sigma (y,t)}}} \right|}  \hfill \\
\end{gathered}
\end{equation}
and thus normalize

\begin{equation}
\tilde p(\tilde t) = 1,\quad\tilde q(\tilde t) = 1,\quad\tilde
\sigma (\tilde y,\tilde t) = \varepsilon  =  \mp 1.
\end{equation}
By an appropriate choice of the functions $\beta(y,t)$ and
$S(y,t)$ we can arrange  to have
$$\tilde{e}(\tilde{y},\tilde{t})=\tilde{h}(\tilde{y},\tilde{t})=0.$$
Finally, equation \eqref{1.1} is reduced to its canonical form

\begin{equation}\label{canon}
\begin{split}
&(u_t+uu_x+u_{xxx})_x+\varepsilon u_{yy}+a(y,t)u_y+b(y,t)u_{xy}\\
&+ c(y,t)u_{xx}+f(y,t)u=0,\quad  \varepsilon=\pm 1.
\end{split}
\end{equation}
With no loss of generality we can restrict our study to symmetries
of eq. \eqref{canon}. All results obtained for eq. \eqref{canon}
can be transformed into results for eq. \eqref{1.1}, using the
transformations \eqref{2.1}. We shall call eq. \eqref{canon} the
"canonical generalized KP equation" (CGKP).

\At{} were used earlier in a very similar manner for a variable coefficient
Korteweg-de-Vries equation \cite{gazeau1, Gungor96}.
\section{Determining equations for the symmetries}
We restrict ourselves to Lie point symmetries. The Lie algebra
of the symmetry group is realized by vector fields of the form

\begin{equation}\label{3.1}
\hat{\mathbf{V}}=\xi\gen x+\eta\gen y+\tau\gen t+\phi\gen u,
\end{equation}
where $\xi$, $\eta$, $\tau$ and $\phi$ are functions of $x,y,t$
and $u$. The algorithm for determining the form of the vector
field $\hat{\mathbf{V}}$ for any given system of differential
equations goes back to S. Lie and is described in any book on the
subject (see e.g. ref. \cite{olver1}). Many computer packages
exist that realize this algorithm. We use the one described in
ref. \cite{Champagne91}. They provide a partially solved set of
determining equations. This is an overdetermined set of linear
partial differential equations for the coefficients $\xi$,
$\eta$, $\tau$ and $\phi$ in eq. \eqref{3.1}.

For eq. \eqref{canon} we obtain 17 equations of which 11 do not
involve the functions $a, b, c$ and $f$. These can be solved to
yield
\begin{equation}\label{3.2}
\begin{aligned}
  \xi (x,y,t,u) &  = \frac{1}
{3}\dot \tau x + \xi _0 (y,t), \\
  \eta (x,y,t,u) &  = \frac{2}
{3}\dot \tau y + \eta _0 (t), \\
  \tau (x,y,t,u) &  = \tau (t), \\
  \phi (x,y,t,u) &  =  - \frac{2}
{3}\dot \tau u + \frac{1}
{3}\ddot \tau x + S(y,t). \\
\end{aligned}
\end{equation}

One of the remaining determining equations can be used to
determine the function $S(y,t)$ to be
\begin{equation}\label{3.3}
S(y,t) =  - \tau c_t  - (\frac{2} {3}\dot \tau y + \eta _0 )c_y +
\xi _{0,t}  + b\xi _{0,y}  - \frac{2} {3}c\dot \tau.
\end{equation}
The remaining determining equations for $\tau(t)$, $\eta(t)$ and
$\xi_0(y,t)$ are

\begin{align}
  & {3\tau a_t  + (2\dot \tau y + 3\eta _0 )a_y  + 2a\dot \tau  = 0}, \hfill\label{3.4}
  \\[.3cm]
  & { - 3\dot \eta _0  - 2y\ddot \tau  + 3\tau b_t  + (2\dot \tau y + 3\eta _0 )b_y
  + b\dot \tau  - 6\varepsilon \xi _{0,y}  = 0}, \hfill \label{3.5}
  \\[.3cm]
  & {\ddot \tau  + 3a\xi _{0,y}  + 3\varepsilon \xi _{0,yy}  = 0}, \hfill\label{3.6}
  \\[.3cm]
  & {f\ddot \tau  = 0}, \hfill \label{3.7} \\[.3cm]
  & {4f\dot \tau  + 3f_t \tau  + f_y (2\dot \tau y + 3\eta _0 ) = 0}, \hfill\label{3.8}
  \\[.3cm]
  & {\dddot \tau  + 3fS + 3aS_y  + 3\varepsilon S_{yy}  = 0},
  \hfill\label{3.9}
 \end{align}
where $S(y,t)$ of eq. \eqref{3.3} should be substituted into eq.
\eqref{3.9}.

It is beyond the scope of this article to perform a complete
analysis of eqs. \eqref{3.3},...,\eqref{3.9} for arbitrary (given)
functions $a$, $b$, $c$ and $f$. Rather, we shall determine the
conditions on these functions that permit the symmetry algebra to
be infinite-dimensional. This will happen when at least one of the
functions $\tau(t)$, $\eta_0(t)$ and $\xi_{0}(y,t)$ remains an
arbitrary function of at least one variable.

From eq. \eqref{3.7} we see immediately that $\tau$ can be
arbitrary only if $f(y,t)$ satisfies $f(y,t)=0$. From eq.
\eqref{3.6} we see $\xi_0(y,t)$ may be an arbitrary function of
$t$, but never of $y$ (we have $\varepsilon=\pm 1$).
\section{Virasoro symmetries of the CGKP equation}
The canonical \gkpe{}  \eqref{canon} will be invariant  under a
transformation group, the Lie algebra of which is isomorphic to a
Virasoro algebra if the function $\tau$ in \eqref{3.2} remains
free. Thus, we are looking for conditions on the coefficients $a,
b, c$ and $f$ that allow equations \eqref{3.4},...\eqref{3.9} to
be solved without imposing any conditions on $\tau(t)$.

From eq. \eqref{3.7} we see that $\tau$ is linear in $t$, unless
we have $f(y,t)\equiv 0$. Once this condition is imposed,
equations \eqref{3.7} and \eqref{3.8} are solved identically. Eq.
\eqref{3.4} leaves $\tau(t)$ free if either we have $a=0$, or
$a=a_0(y+\lambda(t))^{-1}$ where $a_0\ne 0$ is a constant and
$\lambda(t)$ is some function of $t$. We investigate the two
cases separately. First let us assume
\begin{equation}\label{4.1}
a=\frac{a_0}{y+\lambda(t)},\quad a_0\ne 0.
\end{equation}
Then we view eq. \eqref{3.4} as an equation for $\eta_0(t)$ and
obtain
\begin{equation}\label{4.2}
\eta_0(t)=\frac{1}{3}(2\lambda \dot{\tau}-3\dot{\lambda}\tau).
\end{equation}
Eq. \eqref{3.6} allows us to determine $\xi_0(y,t)$ in terms of
$\tau(t)$. Three possibilities occur:

\noindent 1.) $a_0\varepsilon\ne \pm 1$

\begin{equation}\label{4.3}
\xi _0  =  - \frac{{\varepsilon \ddot \tau (y + \lambda )^2 }}
{{6(1 + a_0 \varepsilon )}} + \frac{{\mu _1 (t)(y + \lambda )^{ -
a_0 \varepsilon  + 1} }} {{1 - a_0 \varepsilon }} + \mu _0 (t).
\end{equation}

\noindent 2.) $a_0\varepsilon=1$

\begin{equation}\label{4.4}
\xi _0  =  - \frac{{\varepsilon \ddot \tau }} {{12}}(y + \lambda
)^2  + \mu _1 (t)\ln (y + \lambda ) + \mu _0 (t).
\end{equation}

\noindent 3.) $a_0\varepsilon=-1$

\begin{equation}\label{4.5}
\xi _0  =  - \frac{{\varepsilon \ddot \tau }} {{12}}(y + \lambda
)^2 [2\ln (y + \lambda ) - 1] + \mu _1 (t) (y + \lambda )^2  +
\mu _0 (t).
\end{equation}
We must now put $\xi_0$ of \eqref{4.3}, \eqref{4.4} or \eqref{4.5}
into eq. \eqref{3.5} and solve the obtained equation for
$\mu_1(t)$. The expression for $\mu_1(t)$ must be independent of
$y$ for all values of $\tau$. Moreover, for $\tau(t)$ to remain
free, there must be no relation between $b(y,t)$ and $\tau(t)$.
These conditions cannot be satisfied for any value of
$a_0\varepsilon$. Hence, if $a(y,t)$ is as in eq. \eqref{4.1} the
generalized KP equation \eqref{canon} does not allow a Virasoro
algebra.

The other case to consider is $a=0$ (in addition to $f=0$). Eq.
\eqref{3.6} is easily solved in this case and we obtain
\begin{equation}\label{4.6}
\xi _0(y,t)  =  - \frac{{\varepsilon \ddot \tau }} {6}y^2  + \mu
_1 (t)y + \mu _0 (t)
\end{equation}
with $\mu_1(t)$ and $\mu_0(t)$  arbitrary. We insert $\xi_0(y,t)$
into eq. \eqref{3.5} and solve for $\mu_1(t)$. This is possible
if and only if we have $b=b(t)$ (no dependence on $y$). We obtain
\begin{equation}\label{4.7}
\mu _1 (t) = \frac{\varepsilon } {6}(b\tau  + 3\dot b\tau  -
3\dot \eta _0 ).
\end{equation}
We put the expression \eqref{3.3} for $S(y,t)$ into eq.
\eqref{3.9} and obtain
\begin{equation}\label{4.8}
 - 2\dot \tau (yc_{yyy}  + 3c_{yy} ) - 3c_{tyy} \tau  + 3\eta _0 c_{yyy}  = 0.
\end{equation}
Eq. \eqref{4.8} restricts the form of $\tau(t)$, unless we have
\begin{equation}\label{4.9}
c(y,t)=c_0(t)+c_1(t)y.
\end{equation}
The result is that for $a=f=0$, $b=b(t)$, $c=c_0(t)+c_1(t)y$ all
equations \eqref{3.4},...,\eqref{3.9} are solved with $\tau(t)$,
$\eta_0(t)$ and $\mu_0(t)$ arbitrary. Let us sum up the result as
two theorems.

\begin{thm}\label{t1}
The canonical \gkpe{} \eqref{canon} allows the Virasoro algebra as
a symmetry algebra if and only if the coefficients satisfy
\begin{equation}\label{4.10}
a=f=0,\quad b=b(t),\quad c=c_0(t)+c_1(t)y.
\end{equation}
\end{thm}
\begin{thm}\label{t2}
The canonical generalized KP equation
\begin{equation}\label{4.11}
(u_t+uu_x+u_{xxx})_x+\varepsilon
u_{yy}+b(t)u_{xy}+[c_0(t)+c_1(t)y]u_{xx}=0
\end{equation}
with $\varepsilon=\pm 1$, and $b(t)$, $c_0(t)$ and $c_1(t)$
arbitrary smooth functions is invariant under an
infinite-dimensional Lie point symmetry group. Its Lie algebra has
a Kac-Moody-Virasoro structure. It is realized by vector fields of
the form
\begin{equation}\label{4.12}
\hat{\mathbf{V}}=T(\tau)+X(\xi)+Y(\eta),
\end{equation}
where $\tau(t)$, $\xi(t)$ and $\eta(t)$ are arbitrary smooth
functions of time and we have
\begin{equation}\label{4.13}
\begin{gathered}
  T(\tau ) = \tau (t)\partial _t  + \frac{1}
{6}[3\varepsilon \dot by\tau  + (2x + \varepsilon by)\dot \tau  - \varepsilon \ddot \tau y^2 ]\partial _x  \hfill \\
   + \frac{2}
{3}\dot \tau \partial _y  + \frac{1}
{6}\{ [ - 6\dot c_0  + 3\varepsilon b\dot b + ( - 6\dot c_1  + 3\varepsilon \ddot b)y]\tau  \hfill \\
   + [ - 4u + \varepsilon b^2  - 4c_0  + 4(\varepsilon \dot b - 2c_1 )y]\dot \tau  + (2x - \varepsilon by)\ddot \tau  \hfill \\
   - \varepsilon y^2 \dddot \tau \} \partial _u,  \hfill \\
\end{gathered}
\end{equation}
\begin{align}
   X(\xi ) &= \xi (t)\partial _x  + \dot \xi (t)\partial _u,  \label{4.14} \\
  Y(\eta )& = \eta (t)\partial _y  - \frac{\varepsilon }
{2}\dot \eta (t)y\partial _x  - \frac{1} {2}[2c_1 \eta
   + \varepsilon b\dot \eta  + \varepsilon y\ddot \eta ]\partial
   _u.\label{4.15}
\end{align}
\end{thm}
The form of eq. \eqref{4.11} and its symmetry algebra
\eqref{4.12},...,\eqref{4.15} suggests that it might be
transformable into the KP equation itself. This is indeed the
case. The transformation
\begin{equation}\label{4.16}
\begin{gathered}
  u(x,y,t) = \tilde u(\tilde x,\tilde y,\tilde t) + (\frac{\varepsilon }
{2}\dot b - c_1 )y - c_0  + \frac{\varepsilon }
{4}b^2,  \hfill \\
  \tilde x = x - \frac{{\varepsilon b}}
{2}y,\;\;\tilde y = y,\;\;\tilde t = t \hfill \\
\end{gathered}
\end{equation}
takes eq. \eqref{4.11} into the KP equation itself, i.e. into eq.
\eqref{4.11} with $b=c_0=c_1=0$. The transformation \eqref{4.16}
also transforms the Lie algebra \eqref{4.12},...,\eqref{4.15} into
the symmetry algebra \cite{david1} of the KP equation.

We have obtained the following result.
\begin{thm}\label{t3}
The GKP equation \eqref{1.1} is invariant under a Lie point
symmetry group, the Lie algebra of which contains a Virasoro
algebra as a subalgebra, if and only if it can be transformed into
the KP equation itself by a point transformation.
\end{thm}
\section{Kac-Moody symmetries of the CGKP \\e\-qua\-tion}
In section 4 we have shown that if the symmetry algebra of the
canonical generalized Kadomtsev-Petviashvili  contains a Virasoro
algebra, then it also contains a Kac-Moody algebra. In this section
we will determine the conditions on the functions $a(y,t)$,
$b(y,t)$, $c(y,t)$ and $f(y,t)$ under which the CGKP equation
allows a Kac-Moody algebra, without allowing a Virasoro one. Thus,
the function $\tau(t)$ will not be free, but $\eta_0(t)$ of eq.
\eqref{3.2} will be free, or $\xi_0(y,t)$ will involve at least one
free function of $t$.
\subsection{The function $\eta_0(t)$ free}
Eq. \eqref{3.4} will relate $\eta_0$ and $a(y,t)$ unless we have
$a_y=0$. Hence we put $a_y=0$. For $a=a(t)\ne 0$ eq. \eqref{3.4}
implies $\tau(t)=\tau_0a^{-3/2}$. Eq. \eqref{3.6} then yields
$$\xi=\xi_1(t)e^{-a\varepsilon y}-\frac{\ddot{\tau}}{3a}y+\xi_0(t).$$
Eq. \eqref{3.5} then provides a relation between $\eta_0(t)$ and
$b(y,t)$. Hence $\eta_0(t)$ is not free. Thus, if $\eta_0(t)$ is to
be a free function, we must have $a(y,t)=0$. Eq. \eqref{3.4} is
satisfied identically. From eq. \eqref{3.6} we have
\begin{equation}\label{5.1}
\xi _0 (y,t) =  - \frac{\varepsilon } {6}\ddot \tau y^2  + \rho
(t)y + \sigma (t).
\end{equation}
Eq. \eqref{3.5} will leave $\eta_0$ free only if we have
\begin{align}
b(y,t)&=b_1(t)y+b_0(t),\label{5.2}\\[.2cm]
\rho(t)&=\frac{\varepsilon}{6}(-3\dot{\eta_0}+3\tau \dot{b}_0+3\eta_0
b_1+b_0 \dot{\tau}),\label{5.3}\\[.2cm]
 {(\tau b_1)}^{.}&=0\label{5.4}.
\end{align}
For $f\ne 0$ we have $\ddot{\tau}=0$ and eq \eqref{3.9} will
relate $\eta(t)$ to $c(y,t)$, $b_1$ and $b_0$. Thus, for
$\eta_0(t)$ to be free, we must have $f(y,t)=0$. Eq. \eqref{3.9}
reduces to
\[
\dddot \tau  + 3\varepsilon S_{yy}  = 0.
\]
This equation is only consistent if we have

\begin{align}
& c(y,t) = c_2 (t)y^2  + c_1 (t)y + c_0 (t), \label{5.5}\\[.3cm]
& \varepsilon b_1 \ddot \tau  + 3\dot c_2 \tau  + 6\dot \tau c_2
= 0.\label{5.6}
\end{align}
The only equation that remains to be solved is eq. \eqref{5.6}.
Both functions $\eta_0(t)$ and $\sigma(t)$ remain free. If we have
$b_1=0$, $c_2=0$, then the function $\tau(t)$ is also free and we
reobtain the entire Kac-Moody-Virasoro algebra of Section 4. The
most general CGKP equation allowing $\eta_0(t)$ to be a free
function is obtained if eq. \eqref{5.6} is solved identically by
putting $\tau=0$. Then $\eta_0(t)$ and $\sigma(t)$ are arbitrary.
Using eq. \eqref{3.2} and the above results we obtain the
following theorem.
\begin{thm}\label{t4}
The equation
\begin{equation}\label{5.7}
\begin{split}
(u_t&+uu_x+u_{xxx})_x+\varepsilon u_{yy}+(b_1(t)y+b_0(t))u_{xy}\\
&+(c_2(t)y^2+c_1(t)y+c_0(t))u_{xx}=0,
\end{split}
\end{equation}
where $\varepsilon=\pm 1$ and $b_0, b_1, c_0, c_1, c_2$ are
arbitrary functions of $t$, is the most general canonical \gkpe{},
invariant under an infinite-dimensional Lie point symmetry group
depending on two arbitrary functions. Its Lie algebra has a
Kac-Moody structure \cite{Goddard86} and is realized by vector
fields of the form

\begin{equation}\label{5.8}
\hat{\mathbf{V}}=X(\xi)+Y(\eta),
\end{equation}
where $\xi(t)$ and $\eta(t)$ are arbitrary smooth  functions of
time and
\begin{align}
X(\xi)&=\xi\gen x+\dot{\xi}\gen u, \label{5.9}\\
\begin{split}\label{5.10}
Y(\eta)&=\eta\gen
y+\frac{\varepsilon}{2}y(-\dot{\eta}+b_1\eta)\gen
x+\curl{[-2c_2\eta\\
&+\frac{\varepsilon}{2}(-\ddot{\eta}+\dot{b}_1\eta+b_1^2\eta)]y
-c_1\eta+\frac{\varepsilon}{2}b_0(-\dot{\eta}+b_1\eta)}\gen u.
\end{split}
\end{align}
\end{thm}
Several comments are in order:

\noindent {\bf 1.} The symmetry algebra of eq. \eqref{5.7} is
larger for special cases of the functions $b_1$, $b_0$, $c_2$,
$c_1$ and $c_0$. Thus, if $c_2=b_1=0$ we recover  the entire
Kac-Moody-Virasoro algebra of Theorem \ref{t2} since eq.
\eqref{5.6} is satisfied identically (i.e. the function $\tau(t)$
is also free). The two other special cases are $b_1=0$, $c_2\ne 0$
and $b_1\ne 0$ (see below).

\noindent {\bf 2.} Eq. \eqref{5.7} can be further simplified by
\at{}. Indeed, let us restrict the transformation \eqref{2.1} to
\begin{equation}\label{5.11}
\begin{gathered}
  u(x,y,t) = \tilde u(\tilde x,\tilde y,\tilde t) + S_1 (t)y + S_0 (t), \hfill \\
  \tilde x = x + \beta _1 (t)y + \beta _0(t) ,\;\;\tilde y = y + \gamma (t),\;\;\tilde t = t. \hfill \\
\end{gathered}
\end{equation}
For any functions $b_1(t)$ and $c_2(t)$ we can choose $S_1, S_0,
\beta_0, \beta_1$ and $\gamma$ to set $b_0, c_1$ and $c_0$ equal
to zero. Thus, with no loss of generality, we can set
\begin{equation}\label{5.12}
b_0(t)=c_1(t)=c_0(t)=0
\end{equation}
in eq. \eqref{5.7}, \eqref{5.9} and \eqref{5.10}.

\noindent {\bf 3.} Let us now consider the cases when eq.
\eqref{5.7} has an additional symmetry.

{\bf Case 1.} $b_1=0, c_2\ne 0$

We assume \eqref{5.12} is already satisfied. From eq. \eqref{5.6}
we have $\tau=\tau_0 c_2^{-1/2}$. The additional symmetry is

\begin{equation}\label{5.13}
\begin{split}
  T &= \frac{1}
{{\sqrt {c_2 } }}\partial _t  + [\frac{1} {3}\frac{d}{dt}\left(
{\frac{1} {{\sqrt {c_2 } }}} \right)  x - \frac{\varepsilon }
{6}\frac{d^2}{dt^2}\left( {\frac{1}
{{\sqrt {c_2 } }}} \right) y^2 ]\partial _x  \hfill \\
&+  \frac{2} {3}\frac{d}{dt}\left( {\frac{1} {{\sqrt {c_2 } }}}
\right) y\partial _y  + \{  - \frac{2} {3}\frac{d}{dt}\left(
{\frac{1} {{\sqrt {c_2 } }}} \right)  u + \frac{1}
{3}\frac{d^2}{dt^2}\left( {\frac{1}
{{\sqrt {c_2 } }}} \right) x \hfill \\
  & - \frac{\varepsilon }
{6}\frac{d^3}{dt^3}\left( {\frac{1}
{{\sqrt {c_2 } }}} \right) y^2 \} \partial _u,  \hfill \\
\end{split}
\end{equation}

{\bf Case 2.} $b_1\ne 0$, $b_0=c_1=c_0=0$

Eq. \eqref{5.4} implies $\tau=\tau_0b^{-1}$ and eq. \eqref{5.6}
provides the constraint

\begin{equation}\label{5.14}
c_2=\frac{1}{3}(\varepsilon \dot{b}_1+k b_1^2)
\end{equation}
where $k$ is a constant. The additional element of the symmetry
algebra in this case is

\begin{equation}
\begin{split}
  T &= \frac{1}
{b_1}\partial _t  + [\frac{1} {3}\frac{d}{dt}\left( {\frac{1}
{{b_1 }}} \right)  x - \frac{\varepsilon }
{6}\frac{d^2}{dt^2}\left( {\frac{1} {{b_1 }}} \right) y^2
]\partial _x  + \frac{2} {3}\frac{d}{dt}\left( {\frac{1}
{{b_1 }}} \right)  y\partial _y  \hfill \\
   &+ [ - \frac{2}
{3}\frac{d}{dt}\left( {\frac{1} {{b_1 }}} \right)  u + \frac{1}
{3}\frac{d^2}{dt^2}\left( {\frac{1} {{b_1 }}} \right) x -
\frac{\varepsilon } {6}\frac{d^3}{dt^3}\left( {\frac{1}
{{b_1 }}} \right) y^2 ]\partial _u . \hfill \\
\end{split}
\end{equation}

\subsection{One free function in symmetry algebra}
We have established that if $\tau(t)$ is free in eq. \eqref{3.2},
then there are three free functions. If $\tau$  is not free, but
$\eta_0(t)$ is, then there are two free functions. Now let
$\tau(t)$ and $\eta_0(t)$ be constrained by the determining
equations, but let some freedom remain in the function
$\xi_0(y,t)$.

First of all we note that if we put

\begin{equation}\label{5.16}
\tau=0, \quad \eta_0=0,\quad \xi_0(y,t)=\xi(t)
\end{equation}
in eq. \eqref{3.2} then eqs. \eqref{3.4},...,\eqref{3.8} are
satisfied identically and eq. \eqref{3.9} reduces to

\begin{equation}\label{5.17}
f\dot{\xi}=0.
\end{equation}
Hence

\begin{equation}\label{5.18}
X(\xi)=\xi(t)\gen x+\dot{\xi}(t)\gen u,
\end{equation}
with $\xi(t)$ arbitrary, generates Lie point symmetries of the
CGKP equation for $f(y,t)=0$ and any functions $a(y,t), b(y,t)$,
and $c(y,t)$.

For $f\ne 0$ we have $\tau=\tau_1 t+\tau_0$ from eq. \eqref{3.7}.
Eq. \eqref{3.6} then determines the $y$ dependence of $\xi_0$.
Two possibilities occur:
\begin{enumerate}
\item[(i)]
\begin{equation}\label{5.19}
a=0,\quad  \xi_0=\rho(t)y+\sigma(t)
\end{equation}
\item[(ii)]
\begin{equation}\label{5.20}
a \ne 0, \quad  {a(y,t) \equiv  - \varepsilon \frac{{A_{yy} }}
{{A_y }},\quad A_{yy} \ne 0 }, \quad {\xi _0  = \sigma _1
(t)A(y,t) + \sigma _2 (t)}.
\end{equation}
\end{enumerate}
We skip the details here and just state that the remaining
equations \eqref{3.5}, \eqref{3.8} and \eqref{3.9} do not allow
any solutions with free functions.

We state this result as a theorem.
\begin{thm}\label{t5}
The CGKP equation \eqref{canon} is invariant under an
infinite-di\-men\-sion\-al Abelian group generated by the vector
field \eqref{5.18} for $f(y,t)=0$ and $a, b, c$ arbitrary.
\end{thm}

Theorems \ref{t2}, \ref{t4} and \ref{t5} sum up all cases when the
symmetry algebra of the CGKP equation is infinite-dimensional.
\section{Applications and conclusions}
We have identified all cases when the generalized KP equation has
an infinite-dimensional symmetry group. Let us now discuss the
implications of this result.
\subsection{Equation with Kac-Moody-Virasoro symmetry algebra}
We have shown that eq. \eqref{4.11} is the most general CGKP
equation invariant under a Kac-Moody-Virasoro group. Moreover, it
can be transformed into the KP equation itself. It follows that
any solution of the KP equation can be transformed into a solution
of eq. \eqref{4.11}. The corresponding transformation will however
not take solitons into solitons. More generally, it will not
preserve boundary conditions. The Lax pair for the KP equation is
of course well known and much studied \cite{Ablowitz91}. The
transformation inverse to \eqref{4.16} will take it into a Lax
pair for eq. \eqref{4.11}. The transformed Lax pair can then be
simplified by a redefinition of the wave function figuring in it.
In turn, this Lax pair can be used to obtain new solutions of eq.
\eqref{4.11}. Work on this problem is in progress but goes beyond
the scope of the present article. We note here that eq.
\eqref{4.11} can also be transformed into the integrable
cylindrical KP equation
\begin{equation}\label{6.1}
(u_t+u
u_x+u_{xxx})_x+\frac{1}{2t}u_x+\frac{\varepsilon}{4t^2}u_{yy}=0,
\end{equation}
or one of its generalizations, which have been studied
extensively \cite{Calogero82, Levi88}.
\subsection{Equation with nonabelian Kac-Moody symmetry algebra}
The symmetry algebra \eqref{5.8} of eq. \eqref{5.7} is
infinite-dimensional and nonabelian. Indeed, we have
\begin{equation}\label{6.2}
[Y(\eta_1),Y(\eta_2)]=X(\xi),\quad
\xi=-\frac{\varepsilon}{2}(\eta_1\dot{\eta}_2-\dot{\eta}_1\eta_2).
\end{equation}
Unless we have $b_1=c_2=0$ equation \eqref{5.7} is not
integrable. We can however apply the method of symmetry reduction
to obtain particular solutions. The operator $X(\xi)$ of eq.
\eqref{5.9} generates the transformations
\begin{equation}\label{6.3}
\tilde{x}=x+\lambda \xi(t),\quad \tilde{y}=y,\quad
\tilde{t}=t,\quad
\tilde{u}(\tilde{x},\tilde{y},\tilde{t})=u(x,y,t)+\lambda
\dot{\xi}(t),
\end{equation}
where $\lambda$ is a group parameter. We see that \eqref{6.3} is
a transformation to a frame moving with an arbitrary acceleration
in the $x$ direction. For $\xi$ constant this is a translation,
for $\xi$ linear in $t$ this is a Galilei transformation. An
invariant solution will have the form
\begin{equation}\label{6.4}
u=\frac{\dot{\xi}}{\xi}x+F(y,t).
\end{equation}
Substituting into eq. \eqref{5.7} we obtain the family of
solutions
\begin{equation}\label{6.5}
u=\frac{\dot{\xi}}{\xi}x-\frac{\varepsilon}{2}
\;\frac{\ddot{\xi}}{\xi} y^2+ \rho(t)y+\sigma(t)
\end{equation}
with $\rho(t)$ and $\sigma(t)$ arbitrary.

The transformation corresponding to the general element
$Y(\eta)+X(\xi)$ with $\eta\ne 0$ is easy to obtain, but more
difficult to interpret. An invariant solution will have the form
\begin{equation}\label{6.6}
\begin{split}
u&=[-c+\frac{\varepsilon}{4}(\dot{b}+b^2-\frac{\ddot{\eta}}{\eta})]y^2
+\frac{\dot{\xi}}{\eta}y+F(z,t)\\
z&=x+\frac{\varepsilon}{4}(-b+\frac{\dot{\eta}}{\eta})y^2-\frac{\xi}{\eta}y.
\end{split}
\end{equation}
We have put $b_1=b,\; c_2=c,\; b_0=c_1=c_0=0$, which can be done with
no loss of generality. We now put $u$ of eq. \eqref{6.6} into eq.
\eqref{5.7} (for $c_1=c_0=b_0=0$) and obtain the reduced equation
\begin{equation}\label{6.7}
(F_t+F F_z+F_{zzz})_z+\varepsilon
\frac{\xi^2}{\eta^2}F_{zz}+\frac{1}{2}(\frac{\dot{\eta}}{\eta}-b)F_z
-2c\varepsilon+\frac{1}{2}(\dot{b}+b^2-\frac{\ddot{\eta}}{\eta})=0.
\end{equation}
In general, eq. \eqref{6.7} is not integrable. Putting
\begin{equation}\label{6.8}
\begin{aligned}
  F(z,t) &  = \tilde F(\tilde z,\tilde t),\;\;\;\tilde z = z + \beta (t),
  \;\;\;\tilde t = t, \\
  \dot \beta (t) &  =  - \varepsilon \frac{{\xi ^2 }}
{{\eta ^2 }} \\
\end{aligned}
\end{equation}
we eliminate the $F_{zz}$ term. Choosing $\dot{\eta}/\eta=b(t)$ we
obtain the equation
\begin{equation}\label{6.9}
(F_t+F F_z+F_{zzz})_z=2\varepsilon c(t),
\end{equation}
an equation that is not integrable (for $c\ne 0$) but that has
been extensively studied \cite{Johnson97, Grimshaw71}. We note
that for $c=0$ \eqref{6.9} reduces to the KdV equation, even
though the reduced equation \eqref{5.7} is not integrable for
$b_1(t)\equiv b(t)\ne 0.$
\subsection{Equation with Abelian Kac-Moody symmetry algebra}
Let us now consider the CGKP equation \eqref{canon} with $f=0$.
It is invariant under the transformations generated by $X(\xi)$
of eq. \eqref{5.18}. The invariant solution has the form
\eqref{6.4} and $F(t,y)$ will satisfy the linear equation
\begin{equation}\label{6.10}
 \varepsilon F_{yy}+a(y,t)F_y+\frac{\ddot{\xi}}{\xi}  = 0.
\end{equation}
Eq. \eqref{6.10} can be solved for any function $a(y,t)$ and the
general solution is
\begin{equation}\label{6.11}
\begin{split}
F(y,t)&=\alpha(t)A(y,t)+\beta(t)+\varepsilon\frac{\ddot{\xi}}{\xi}
\Bigl[\int_{y_0}^y\frac{A(y',t)}{A_{y'}(y',t)}\,dy'\\
&-\Bigl(\int_{y_0}^y\frac{1}{A_{y'}(y',t)}\,dy'\Bigr)A(y,t)\Bigr],
\end{split}
\end{equation}
where $\alpha(t)$, $\beta(t)$ (and also $\xi(t)$) are arbitrary
functions of $t$. We have put
\begin{equation}\label{6.12}
A(y,t)=\int_{y_0}^y[\exp\bigl(-\varepsilon\int_{y_0}^{y'}
a(y'',t)dy''\bigr)]dy',
\end{equation}
i.e. $A(y,t)$ is a particular solution of \eqref{6.10} for
$\ddot{\xi}=0$ (the homogeneous equation).

For $a(y,t)=0$ the solution \eqref{6.11} reduces to \eqref{6.5},
as it should.

One situation in which solutions of the type \eqref{6.4} could be
relevant is that of water propagating parallel to a beach. The
origin $y=0$ of the $y$ axis would be somewhere in the deep ocean
and $y=y_{\max}$ would be at the shore. The function $a(y,t)$
could be very large for $y$ small, then decrease towards the
shore. An example of such a function and the corresponding
solution are
\begin{equation}\label{6.13}
a(y,t)=-\varepsilon\frac{m(t)}{y},\quad
F(y,t)=A(t)y^{m+1}+\frac{\dot{\xi}}{2\varepsilon\xi(m-1)}y^2+C(t),
\end{equation}
where $m(t)$, $A(t)$ and $C(t)$ are arbitrary functions.

\subsection{Comments and Outlook}
The most ubiquitous symmetry of the generalized KP equation is the
transformation \eqref{6.3} to an arbitrary frame moving in the $x$
direction. Its presence only requires the coefficient $f(y,t)$ in
eq. \eqref{1.1} (or \eqref{canon}) to be $f(y,t)\equiv 0$.
Invariance of a solution under such a general transformation is
very restrictive and leads to solutions that are at most linear in
the variable $x$ and have a prescribed $y$ dependence (see
solutions \eqref{6.5}, \eqref{6.11}, \eqref{6.13}). Such solutions
may actually physically be meaningful under special conditions.
The wave crests $u(x,y,t)=\text{const.}$ for fixed $t$ have a
parabolic shape in the case of eq. \eqref{6.5}. Such "horse-shoe"
shaped waves coming out of straits are observed. Since $u(x,y,t)$
grows linearly with $x$, such solutions can only be physical for a
finite range of values of $x$. After that, presumably the
conditions under which the GKP was derived no longer hold and the
solution breaks, or acquires a different form.

The transformations generated by $Y(\eta)$ leave a more restricted
class of GKP equations invariant, those of eq. \eqref{5.7}. The
invariant solutions have the form \eqref{6.6} and they are much
more general and realistic than the solutions \eqref{6.11}. They
are also much harder to pin down, since it is still necessary to
solve eq. \eqref{6.7} or \eqref{6.9}. For general $c(t)$ this is
difficult, but for $c(t)=0$ this is just the KdV equation, for
arbitrary $b(t)$, as long as we choose $\dot{\eta}/\eta=b(t)$. Any
solution of the KdV equation, in particular soliton, or
multisoliton solutions will, via eq. \eqref{6.6}, provide $y$
dependent solutions of the corresponding GKP equation.

According to our opinion, the most important and interesting
remaining question is: What can we do with the integrable CGKP
\eqref{4.11}? This merits a separate investigation, using the
tools of soliton theory: the inverse spectral transform and
B\"acklund transformations. Both are known for the KP equation
\cite{Levi81, David86, Ablowitz91}. We plan to adapt them to eq.
\eqref{4.11} and to use them to obtain multisoliton and other
physically important solution of this equation. Simple solutions
like \eqref{6.5} can then serve as input solutions into B\"acklund
transformations.

\subsection*{Acknowledgements} Most of the results presented in this
article were obtained while F.G. was visiting the Centre de
Recherches Mathematiqu\'es, Universit\'e de Montr\'eal. He thanks
the CRM for its hospitality and \.{I}stanbul Technical University for
a fellowship that made his visit possible. The final version was
written while P.W. was visiting the Department of Applied
Mathematics of the University of New South Wales. P.W. thanks this
Department and  Prof. C. Rogers for hospitality and support. The
authors thank  Prof. D.Levi for illuminating discussions. The
research of P.W. is partly supported  by research grants from
NSERC of Canada and FCAR du Quebec.

\begin{thebibliography}{10}

\bibitem{David87}
D.~David, D.~Levi, and P.~Winternitz.
\newblock Integrable nonlinear equations for water waves in straits of varying
  depth and width.
\newblock {\em Stud. Appl. Math.}, 76:133--168, 1987.

\bibitem{David89}
D.~David, D.~Levi, and P.~Winternitz.
\newblock Solitons in shallow seas of variable depth and in marine starits.
\newblock {\em Stud. Appl. Math.}, 80:1--23, 1989.

\bibitem{Johnson97}
R.S. Johnson.
\newblock {\em A Modern Introduction to the Mathematical Theory of Water
  Waves}.
\newblock Cambridge University Press, Cambridge, 1997.

\bibitem{Ablowitz81}
M.J. Ablowitz and H.~Segur.
\newblock {\em Solitons and the Inverse Scattering Transform}.
\newblock SIAM, Philadelphia, 1981.

\bibitem{Ablowitz91}
M.J. Ablowitz and P.A. Clarkson.
\newblock {\em Solitons, Nonlinear Evolution Equations and Inverse Scattering}.
\newblock Cambridge University Press, Cambridge, 1991.

\bibitem{Fu84}
L.L. Fu and B.~Holt.
\newblock Internal waves in the gulf of California: Observations from a
  spaceborne radar.
\newblock {\em J. Geophys. Res.}, 89:2053--2060, 1984.

\bibitem{Lacombe82}
H.~Lacombe and C.~Richez.
\newblock {\em The Regime of the Strait of Gibralter}.
\newblock Elsevier, Amsterdam, 1982.

\bibitem{david1}
D.~David, N.~Kamran, D.~Levi, and P.~Winternitz.
\newblock Subalgebras of loop algebras and symmetries of the
  {Kadomtsev-Petviashvili} equation.
\newblock {\em Phys. Rev. Let.}, 55(20):2111--2113, 1985.

\bibitem{Levi88}
D.~Levi and P.~Winternitz.
\newblock The cylindrical Kadomtsev-Petviashvili equation, its
  Kac-Moody-Virasoro algebra and relation to the {KP} equation.
\newblock {\em Phys. Lett. A}, 129:165--167, 1988.

\bibitem{Champagne88}
B.~Champagne and P.~Winternitz.
\newblock On the infinite dimensional symmetry group of the Davey-Stewartson
  equation.
\newblock {\em J. Math. Phys.}, 29:1--8, 1988.

\bibitem{Winternitz88}
P.~Winternitz.
\newblock Kac-Moody-Virasoro symmetries of integrable nonlinear partial
  differential equations.
\newblock In D.~Levi and P.~Winternitz, editors, {\em Symmetries and Nonlinear
  Phenomena}, Singapore, 1988. World Scientific.

\bibitem{Orlov97}
A.Yu. Orlov and P.~Winternitz.
\newblock Algebra of pseudodifferential operators and symmetries of equations
  in the Kadomtsev-Petviashvili hierarchy.
\newblock {\em J. Math. Phys.}, 38:4644--4674, 1997.

\bibitem{Orlov86}
A.Yu. Orlov and E.I. Schulman.
\newblock Additional symmetries for integrable eqautions and conformal algebra
  representations.
\newblock {\em Lett. Math. Phys.}, 12:171--179, 1986.

\bibitem{Clarkson90}
P.A. Clarkson.
\newblock Painlev\'e analysis and the complete integrability of a generalized
  variable-coefficient Kadomtsev-Petviashvili equation.
\newblock {\em IMA J. Appl. Math.}, 44:27--53, 1990.

\bibitem{Ablowitz78}
M.J. Ablowitz, A.~Ramani, and H.~Segur.
\newblock Nonlinear evolution equations and ordinary differential equations of
  the Painlev\'e type.
\newblock {\em Lett. Nuovo Cim.}, 23:333--338, 1978.

\bibitem{Weiss83}
J.~Weiss, M.~Tabor, and G.~Carnevale.
\newblock The Painlev\'e property for partial differential equations.
\newblock {\em J. Math. Phys.}, 24:522--526, 1983.

\bibitem{Conte99}
R.~Conte, editor.
\newblock {\em The Painlev\'e Property, One Century Later, {CRM} Series in
  Mathematical Physics}, NewYork, 1999. Springer.

\bibitem{gazeau1}
J.P. Gazeau and P.~Winternitz.
\newblock Symmetries of variable coefficient Korteweg-de Vries equations.
\newblock {\em J. Math. Phys.}, 33(12):4087--4102, 1992.

\bibitem{Gungor96}
F.~G{\"u}ng{\"o}r, M.~Sanielevici, and P.~Winternitz.
\newblock On the integrability properties of variable coefficient Korteweg-de
  Vries equations.
\newblock {\em Can. J. Phys.}, 74:676--684, 1996.

\bibitem{olver1}
P.J. Olver.
\newblock {\em Applications of Lie Groups to Differential Equations}.
\newblock Springer, New York, 1986.

\bibitem{Champagne91}
B.~Champagne, W.~Hereman, and P.~Winternitz.
\newblock The computer calculation of Lie point symetries of large systems of
  differential equations.
\newblock {\em Comp. Phys. Comm.}, 66:319--340, 1991.

\bibitem{Goddard86}
P.~Goddard and D.~Olive.
\newblock Kac-Moody-Virasoro algebras in relation to quantum physics.
\newblock {\em Int. J. Mod. Phys.}, A1:303--414, 1986.

\bibitem{Calogero82}
F.~Calogero and A.~Degasperis.
\newblock {\em Spectral Transform and Solitons, Vol.1}
\newblock North Holland, Amsterdam, 1982.

\bibitem{Grimshaw71}
R.~Grimshaw.
\newblock The solitary wave in water of variable depth.
\newblock {\em J. Fluid Mech.}, 46:611--622, 1971.

\bibitem{Levi81}
D. Levi, L. Pilloni and P. Santini.
\newblock B\"acklund transformations for nonlinear evolution
equations in 2+1 dimensions.
\newblock {\em Phys Lett. A}, 81:419--423, 1981.

\bibitem{David86}
D. David, D. Levi, and P. Winternitz.
\newblock B\"acklund transformations and the
infinite dimensional symmetry group of the Kadomtsev-Petviashvili
equation.
\newblock {\em Phys Lett. A}, 118:390--394, 1986.


\end{thebibliography}

\end{document}